\documentstyle[sprocl,definitions,psfig]{article}
\begin{document}

\mitlabels
\ohnelabels
\mitkurzform
\ohnekurzform

\bibliographystyle{unsrt}    

\def\Journal#1#2#3#4{{#1} {\bf #2}, #3 (#4)}

\def\NCA{\em Nuovo Cimento}
\def\NIM{\em Nucl. Instrum. Methods}
\def\NIMA{{\em Nucl. Instrum. Methods} A}
\def\NPB{{\em Nucl. Phys.} B}
\def\PLB{{\em Phys. Lett.}  B}
\def\PRL{\em Phys. Rev. Lett.}
\def\PRD{{\em Phys. Rev.} D}
\def\ZPC{{\em Z. Phys.} C}
\def\st{\scriptstyle}
\def\sst{\scriptscriptstyle}
\def\mco{\multicolumn}
\def\epp{\epsilon^{\prime}}
\def\vep{\varepsilon}
\def\ra{\rightarrow}
\def\ppg{\pi^+\pi^-\gamma}
\def\vp{{\bf p}}
\def\ko{K^0}
\def\kb{\bar{K^0}}
\def\al{\alpha}
\def\ab{\bar{\alpha}}
\def\CPbar{\hbox{{\rm CP}\hskip-1.80em{/}}}
%
\begin{flushright}Imperial/TP/96-97/58\\[12ex]
\end{flushright}

\title{On the Finite Temperature Phase Transition of Scalar
QED
\footnote{Talk given at the E\"otv\"os Conference in Science: Strong and
     Electroweak Matter, Eger, Hungary, 21-25 May 1997, to be
published in the proceedings.}
}

\author{Daniel F. Litim\footnote{E-Mail: D.Litim@ic.ac.uk}}

\address{ Theoretical Physics Group, Blackett Laboratory\\
        Prince Consort Road, Imperial College,
        London SW7 2BZ, U.K.}


\maketitle\abstracts{We analyse the finite temperature phase transition
of scalar QED in euclidean spacetime. Analytical solutions of
approximations to a Wilsonian renormalisation group equation are
discussed.  Special emphasis is put on the
discussion of the $3d$ running of the Abelian charge and its effects
on thermodynamical quantities. An upper bound for the range of
first-order phase transitions is given. 
Our results are compared to
those of resummed perturbation theory.}
\section{Introduction}
The calculation of thermodynamical quantities related to a
first-order phase transition necessitates the knowledge of the
effective potential at the critical temperature. The latter can be obtained in two steps. 
Step I integrates-out the so-called heavy and super-heavy modes, i.e.~the non-zero Matsubara frequencies modes of all fields, as well as the 
Debye mode. This reduces the $4d$ problem to a purely $3d$
one. Integrating-out the remaining fluctuations 
is performed in Step II. In the
parameter range of interest, Step I can be performed
perturbatively.\cite{Kajantie94a} Step II is more
involved. Monte-Carlo techniques\cite{MonteCarlo} have been used for
large Higgs boson mass $m_{\rm H}$
as perturbation theory\cite{32}$^{\!-\,}$\cite{HebeckerDiss} appears to
be applicable only for small $m_{\rm H}$.  However, the $3d$ gauge
coupling $\bar e_3$ 
is  a dimensionful quantity and displays a non-trivial scale
dependence which is not accounted for within the standard perturbative
approach.\cite{Litim95a}  The aim of the present note is to give an
alternative treatment of Step II  clarifying
the impact of a running gauge coupling on thermodynamical quantities at the
example of scalar QED.

\section{Flow equations}
In order to quantify the effect of the running gauge coupling, we will
use a Wilsonian renormalisation group method based upon the effective
average action.\cite{ReuterWetterich93} It gives a prescription about
how an effective action changes with  scale $k$ when fluctuations
with momenta $p$ have been integrated out down to $p\approx k$. As a
result, one 
obtains a coupled set of flow equations w.r.t.~$k$ for both the
$3d$ Abelian charge $\bar e_3$ and the effective average potential
$U_k$. The conventional effective action is obtained in the limit $k \to 0$. 
For simplicity, we will neglect the contributions from the scalar
fluctuations to the scale-dependence of the potential. They are known
to be subleading 
compared to those of the gauge field, as long as $m_{\rm H}$ is
smaller than the photon mass $M$. Furthermore, we will employ a sharp
cut-off regularisation throughout. The (in-)dependence of physical
quantities on the regularisation scheme employed is discussed
elsewhere.\cite{Litim97} Under the above assumptions, the flow
equations for $e^2_3=\eb2_3/k$ and $U_k$ read
\beq
k\0{d \e2_3}{dk}=-\e2_3(1 - \eta_F),\quad k\0{dU_k}{dk}= \0{k^3}{2\pi}
\ln\left(1+\0{2\e2_3(k)\rb}{k}\right). 
\eeq[flowU]
The anomalous dimension $\eta_F$
can be approximated by
$\eta_F=e_3^2/\es2$, with $\es2$ denoting an {\it effective} fixed
point for the dimensionless gauge
coupling.\cite{ReuterWetterich93}$^{\!,\,}$\cite{Litim95a}   For vanishing
scalar and gauge field mass  we find $\es2\approx 6\pi^2$ in agreement
with the $\epsilon$-expansion. However, the effective fixed point
value $\es2$ can be different for non-vanishing masses, which is why
we will keep $\es2$ as a free parameter. The running gauge coupling
follows from the above as  
\beq
\e2_3(k)=\0{\es2}{1+k/\ktr},\quad  \ \ktr= \0{ \La
e^2_3(\La)}{\es2-e^2_3(\La)}. 
\eeq[e2k]
The initial scale $\La$ is proportional to temperature and will be
determined below. Two comments are in order. Firstly, note the appearance of a
characteristic {\it transition scale}  
$\ktr$, describing the  cross-over between the Gaussian and the Abelian
fixed point. For
$k> \ktr$ the running is very slow and dominated by the Gaussian fixed
point, $\eb2_3\sim const.$, whereas for $k<\ktr$ the running
becomes strongly linear and the Abelian fixed point governs the
scale dependence, $\eb2_3(k)\sim k$. Secondly, the limit
$\es2\to\infty$ corresponds formally to neglecting the running of
$\eb2_3$ throughout, because
$\eb2_3(k)=\La e^2_3(\La)=const.$ in this case.

The initial conditions to \Eq{flowU}, obtained through Step I, read at the
scale of dimensional reduction $\La = \xi T$
\beqa
&&U_{\La}(\rb) =
\left(A\ T^2-\012m_{\rm H}^2\right)\rb+\012 B\ T\rb^2,\quad
\eb2_3(\La)= e^2 T,\\[1ex]
&&A ={\e2}/4+{\la}/6-{\e3}/{4\sqrt{3}\pi},\quad B=
{\e4}/{4\pi^2}-{2\e3}/{\sqrt{3}\pi}+\la \ . 
\eeqa[AB]
Here, $\la$ and  $e$ denote resp.~the zero temperature quartic scalar/gauge
field coupling. 
Finally, the effective average potential for scales $k$ smaller than $\La$
follows (with
$\rb=\varphi^*\varphi$) as  
\beq
U_k(\rb) = U_\La(\rb) + \De_k(\rb)\ ,\quad \De_k(\rb)=\int^k_\La dk
\0{k^2}{2\pi}\ln\left(1+\0{2\e2_3(k)\rb}{k}\right)\ . 
\eeq[Uformal]
The integral in \Eq{Uformal} can be solved analytically. Its solution
is discussed in the following section.

\section{Criticality}
We are now interested in the characteristics of the first order phase
transition, using $e=0.3$ and  $M=80.6$ GeV as initial
parameters. The latter is 
related to the coupling $\la$ through $\la/e^2 = m_{\rm H}^2/M^2$.
The physical quantities that characterise a first-order phase
transition  are defined at the
critical temperature $T_c$, when the
potential has two degenerate 
minima at $\rb=0$ and a non-trivial one at $\rb=\rb_0\neq 0$. Using
$x={\rb_0}/{T}$ and $\t{\De}=\De/T^3$, the  conditions for a
degenerate potential read
\beq
\0{m_{\rm H}^2}{T_c^2}-2A\ =\ \02x\left[2\t{\De}(x)-x\t{\De}'(x)\right],\quad
B\ =\ \02{x^2}\left[\t{\De}(x)-x\t{\De}'(x)\right] \ .
\eeq[crit]
\begin{figure}[t]
\begin{center}
\rule{12cm}{0mm}
\unitlength0.001\hsize
\begin{picture}(1000,500)
\put(200,50){ $m_{\rm H}$}
\put(150,350){\Large $\xi=\La/T$}
\put(750,50){ $\varphi/v$}
\put(650,350){\Large $10^7\ U_{\rm crit}/v^4$}
\put(600,480){\footnotesize
\begin{tabular}{ll}
$\large{}_{e_\star=\sqrt{6} \pi}$&$  \put(0,0){\line(70,0){70}}${}\\[-.5ex]
$\large{}_{e^4, \la^2}$&$  \multiput(0,0)(20,0){4}{\line(10,0){10}}${}\\[-.5ex]
$\large{}_{e^3, \la^{3/2}}$&$  \multiput(0,0)(10,0){7}{\line(5,0){5}}$
\end{tabular}}

\psfig{file=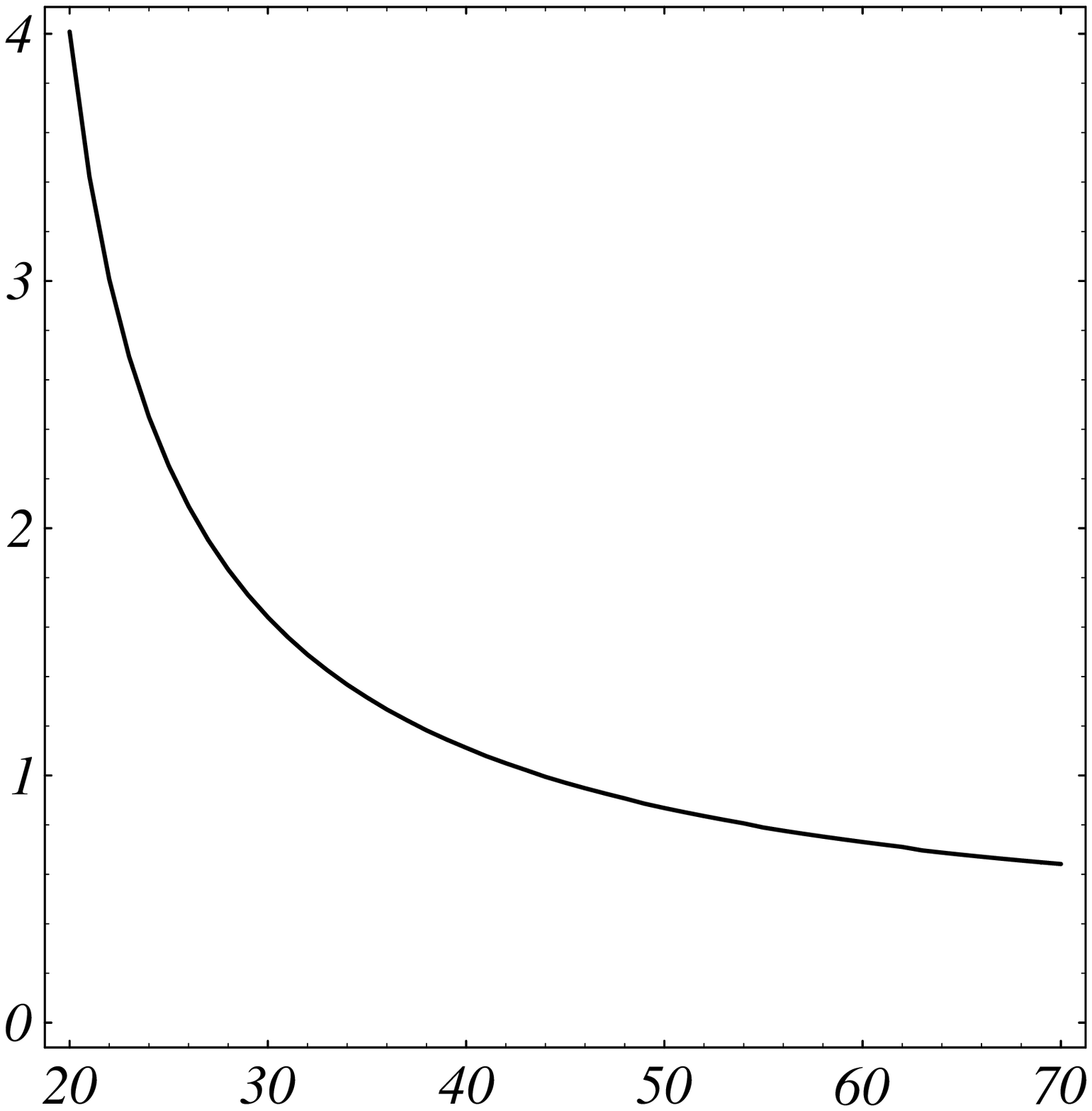,width=.45\hsize}
\hskip.09\hsize\psfig{file=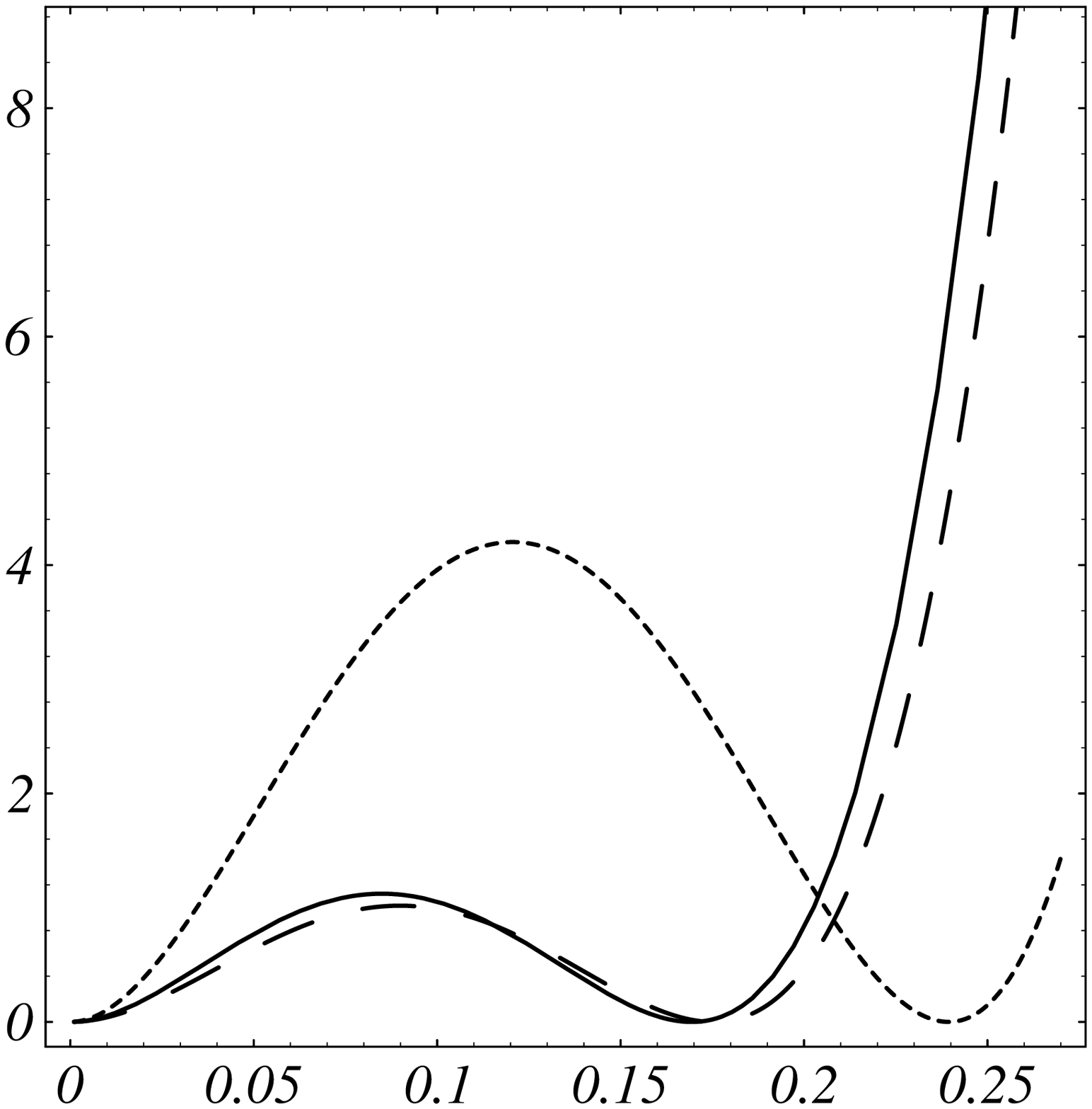,width=.45\hsize}
\end{picture}
\vskip-.5cm \begin{minipage}{.45\hsize}
\caption{\small  The $3d$ matching scale as a function
of the Higgs boson mass.}
\end{minipage} 
\hfill\begin{minipage}{.45\hsize}
\caption{\small  The crit.~potential in different approximations, $m_{\rm H}=38$
GeV .}\end{minipage} 
\end{center}
\end{figure}
\begin{figure}[ht]
\begin{center}
\rule{12cm}{0mm}
\unitlength0.001\hsize
\begin{picture}(1000,500)
\put(700,50){ $\mbox{\small{$ \frac{1}{2} $}}\ln (\es2/6 \pi^2)$}
\put(750,350){\Large ${T_c}/{m_{\rm H}}$}
\put(200,50){ $m^2_{\rm H}/M^2$}
\put(250,350){\Large $\varphi_0/v$}
\put(200,480){\footnotesize
\begin{tabular}{ll}
$\large{}_{e_\star= 100}$&$  \put(0,0){\line(70,0){70}}${}\\[-.5ex]
$\large{}_{e^4, \la^2}$&$  \multiput(0,0)(20,0){4}{\line(10,0){10}}${}\\[-.5ex]
$\large{}_{e^3, \la^{3/2}}$&$  \multiput(0,0)(10,0){7}{\line(5,0){5}}$
\end{tabular}}
\psfig{file=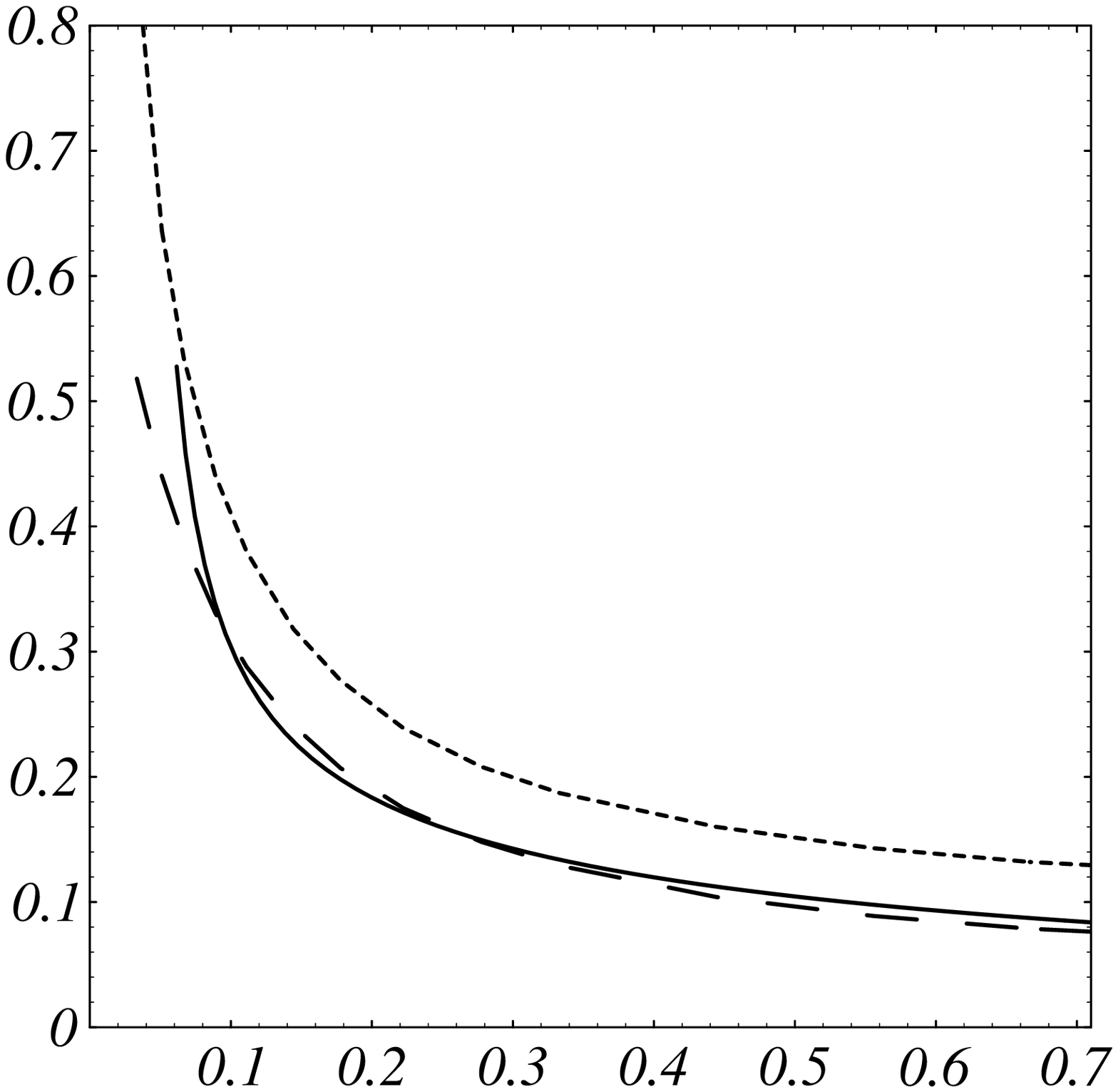,width=.47\hsize} \hskip.04\hsize\psfig{file=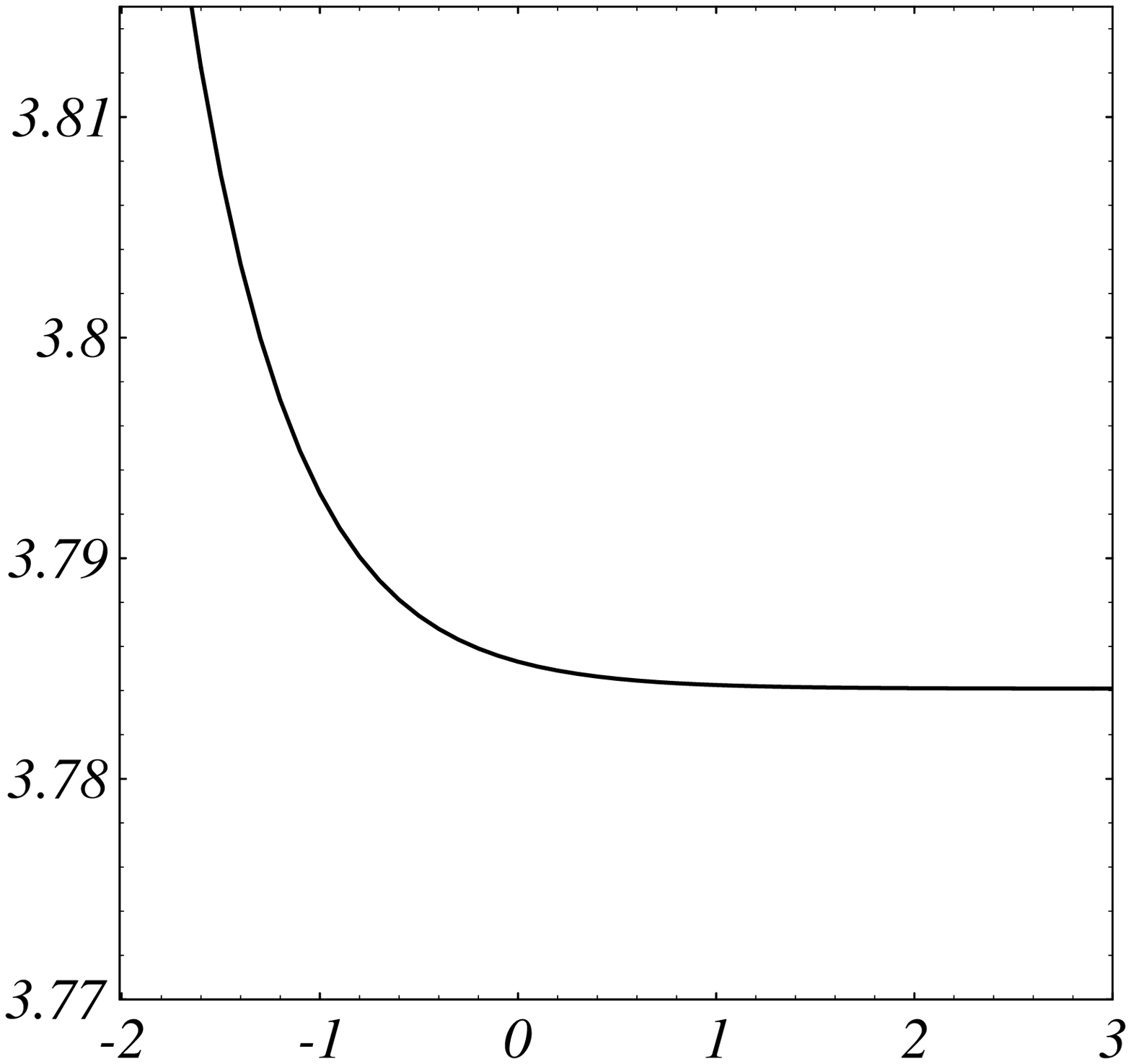,width=.47\hsize}\end{picture}
\begin{minipage}{.45\hsize}
\vskip-.5cm \caption{\small  The v.e.v.~$\varphi_0 $ as a function
of the Higgs boson mass. $v$ denotes the $T=0 $ v.e.v.~of $\varphi$.}
\end{minipage} 
\hfill\begin{minipage}{.45\hsize}
\vskip-.5cm \caption{\small  The critical temperature as a function of
$\es2$ at $m_{\rm H}=38$ GeV .}
\end{minipage} 
\end{center}
\end{figure}
Note that, for $x>0$, the r.h.sides are  positive, finite,
monotonically decreasing and 
vanishing for $x\to\infty$. They reach their respective  maximum at
$x=0$. $B$, as given by  \Eq{AB} and fixed
through the $4d$ parameters of the theory, is positive in 
the domain under consideration. It follows that a solution to
\Eq{crit} is unique (if it 
exists). There exists no solution for too large
values of $B$. Thus, 
\beq 
\0{m_{\rm H}^2}{M^2}\approx \0{8\es2}{3\pi^2}
\eeq[mh-crit]
is an upper bound for the scalar mass. For larger $m_{\rm H}$ the phase
transition ceases to be first order. This follows from the very
existence of an effective fixed point for the Abelian charge and is
in contrast to the perturbative analysis, since a non-running of the
gauge coupling 
($\es2\to\infty$) would predict a first order phase transition for all
$m_{\rm H}$. We interpret this upper limit as a sign for the existence
of a tricritical fixed point that marks the endpoint of a line of
first order phase transitions. For too large values of $\es2$, however, 
this argument is no longer valid as \Eq{mh-crit}  lies outside the
domain of validity. 

Our results are displayed in the figures 1 -- 4. Fig.~1 shows the
optimised matching parameter $\xi$  as a function of the Higgs boson
mass. It obtains as the unique solution to the minimum sensitivity
condition $d\rb_0(\xi)/d\xi=0$, and ensures that physical quantities
do not depend (to leading order) on the details of the matching
itself. The magnitude of $\xi$ for small $m_{\rm H}$ indicates that
the two-step matching  is not reliable in this region. For larger Higgs boson mass, $\xi$ becomes nearly independent of $m_{\rm H}$.

Figs.~2 and 3  compare our results with those from resummed
perturbation theory.\cite{32}$^{\!-\,}$\cite{HebeckerDiss} The critical
potential at $m_{\rm H}=38$ GeV for $e_\star=\sqrt{6}\pi$ (Fig.~2) comes
quite close to the two-loop perturbative result.\cite{Hebecker93} This
indicates that the gauge field  fluctuations do indeed give the
largest contribution to the running potential. We checked also that
the leading contributions of the scalar fluctuations give only a small
effect. The
same holds true for the v.e.v.~as a function of $m_{\rm H}$
(Fig.~3). We wish to emphasize that the latent heat  $L=T\partial_T
U(\rb_0,T_c)$ can be obtained directly from fig.~3, because any
solution to \Eq{crit} automatically fulfills the Clausius-Clapeyron
equation $L=\rb_0 \ m^2$. This feature was also observed within a
gauge-invariant perturbative calculation, but not within the standard
perturbative approach.\cite{HebeckerDiss}

The dependence of the critical temperature on the fixed point value
can be read off from fig.~4.  Note that  $T_c$ remains independent of
$\es2$ for $\es2>6 \pi^2$. Even for small values of $\es2$, $T_c$
increases only slowly. This is the region where  the running of the
gauge coupling becomes important due to a sufficiently small value of
the fixed point. For large $\es2$ and small $m_{\rm H}$ the scale
$\k-{tr}$ is much smaller then the characteristic scales relevant for
the first-order phase transition. The scale dependence of the gauge
coupling is therefore of no importance in this domain and  its
neglection seems to be well justified.

To conclude, we have seen that the results of two-loop perturbation
theory can be obtained within a simple and systematic approximation to a Wilsonian
RG.  The very existence of a (partial) fixed point for the gauge
coupling seems to imply the existence of a tricritical fixed
point. Quantitatively, the effects from the running of the gauge coupling can be neglected for small
$m_{\rm H}$. However, they are getting stronger for larger $m_{\rm H}$
and smaller $\es2$. With $\es2$ near or above its perturbative
value $\approx 6 \pi^2$ the effects
are of the order of percents or smaller.

\section*{Acknowledgements}
I would like to thank F.~Freire for collaboration and enjoyable
discussions, and A.~Hebecker for providing his data in figs.~2,3.
This work was supported in part by the European Comission under the Human Capital and Mobility Programme, contract number CHRX-CT94-0423.



\begin{thebibliography}{99}
\def\BOOK#1#2#3#4{#1, {\sc #2} #3, #4}
\def\PRA#1#2#3#4#5{ #1, {\it } Phys. Rev.~{\bf A #3} (19#4) #5}
\def\PRB#1#2#3#4#5{#1, {\it } Phys. Rev.~{\bf B #3} (19#4) #5}
\def\PRL#1#2#3#4#5{#1, {\it } Phys. Rev.~Lett.~{\bf #3} (19#4) #5}
\def\PRC#1#2#3#4#5{#1, {\it } Phys. Rev.~{\bf C #3}  (19#4) #5}
\def\PRD#1#2#3#4#5{#1, {\it } Phys. Rev.~{\bf D #3} (19#4) #5}
\def\PRE#1#2#3#4#5{#1, {\it } Phys. Rev.~{\bf E #3} (19#4) #5}
\def\PRep#1#2#3#4#5{#1, {\it } Phys. Rep.~{\bf  #3} (19#4) #5}
\def\NPB#1#2#3#4#5{#1, {\it } Nucl. Phys.~{\bf B #3} (19#4) #5}
\def\PLB#1#2#3#4#5{#1, {\it } Phys. Lett.~{\bf B #3} (19#4) #5}
\def\PTP#1#2#3#4#5{#1, {\it } Prog. Theor.~Phys.~{\bf B #3} (19#4) #5}
\def\SSC#1#2#3#4#5{#1, {\it } Solid State Comm.~{\bf  #3} (19#4) #5}
\def\EPL#1#2#3#4#5{#1, {\it } Europhys. Lett.~{\bf #3} (19#4) #5}
\def\JCP#1#2#3#4#5{#1, {\it } J.~Phys. (Paris) {\bf  #3} (19#4) #5}
\def\JPA#1#2#3#4#5{#1, {\it } J.~Phys. {\bf A  #3} (19#4) #5}
\def\JPB#1#2#3#4#5{#1, {\it } J.~Phys. {\bf B  #3} (19#4) #5}
\def\JPC#1#2#3#4#5{#1, {\it } J.~Phys. {\bf C  #3} (19#4) #5}
\def\ZPC#1#2#3#4#5{#1, {\it } Z.~Phys. {\bf C  #3} (19#4) #5}
\def\JETP#1#2#3#4#5{#1, {\it } Soviet Physics JETP Lett.~{\bf #3} (19#4) #5}
\def\MPLA#1#2#3#4#5{#1, {\it } Mod.~Phys. Lett.~{\bf A  #3} (19#4) #5}
\def\PA#1#2#3#4#5{#1, {\it } Physica {\bf A  #3} (19#4) #5}
\def\PS#1#2#3#4#5{#1, {\it } Physics {\bf   #3} (19#4) #5}
\def\AP#1#2#3#4#5{#1, {\it } Ann. Phys. {\bf  #3} (19#4) #5 }
\def\IJMPA#1#2#3#4#5{#1, {\it } Int.~J. Mod. Phys.~ {\bf A  #3} (19#4) #5}
\def\LNC#1#2#3#4#5{#1, {\it } Lett.~Nuevo Cimento {\bf   #3} (19#4) #5}
\def\PPR#1#2#3{#1,  {\it }  #3}


\bibitem{Kajantie94a}\NPB{K.~Farakos, K.~Kajantie, K.~Rummukainen,
M.~Shaposhnikov}{}{425}{94}{67}. 
\bibitem{MonteCarlo}\PPR{M.~Karjalainen, J.~Peisa}{Dimensionally reduced
U(1)+Higgs theory in the broken phase}{hep-lat/9607023};
 \PPR{P.~Dimouploulos, K.~Farakos, G.~Koutsoumbas}{Three-dimensional lattice
U(1) gauge-Higgs model}{hep-lat/9703004}; \PPR{K.~Kajantie,
M.~Karjalainen, M.~Laine, J.~Peisa}{}{}{cond-mat/9704056}.
\bibitem{32}\NPB{W. Buchm\"uller, T. Helbig, D. Walliser}{}{407}{93}{387}. 
\bibitem{Hebecker93}\ZPC{A.~Hebecker}{Finite Temperature Effective
Potential for the Abelian Higgs Model to the Order
$\e4,\la^2$}{60}{93}{271}.
\bibitem{HebeckerDiss}A.~Hebecker, Ph.-D. Thesis (hep-ph/9506418) and references therein.
\bibitem{ReuterWetterich93}\NPB{M.~Reuter, C.~Wetterich}{}{391}{93}{91}.
\bibitem{Litim95a}\PRD{B.~Bergerhoff, F.~Freire, D.F.~Litim, S.~Lola,
C.~Wetterich}{Phase Diagram of Superconductors from Non Perturbative
Flow Equations}{53}{96}{5734}; \IJMPA{B.~Bergerhoff, D.F.~Litim, S.~Lola,
C.~Wetterich}{Phase Transition of $N$-Component
Superconductors}{11}{96}{4273}.
\bibitem{Litim97}\PLB{D.F.~Litim}{Scheme Independence and First Order
Phase Transitions from Exact Renormalization Group
Equations}{393}{97}{103}; F.~Freire, D.F.~Litim, under completion.
\end{thebibliography}
\end{document}